\documentclass[12pt,a4paper,dvips]{article}
\usepackage{a4p}
\usepackage{cite,mcite} 
\usepackage{graphicx}
\usepackage{physics}
\usepackage{l3_title,ifthen,Lep}
\usepackage{epsfig,float}
\journalname{Phys. Lett. B}
\preprint{2003-085}
\date{December 9, 2003}
\def\ra{\rightarrow }

\def\epem{\mbox{e}^+\mbox{e}^- }

\def\wgg{W_{\gamma \gamma}}
\def\ra{\rightarrow } 
\def\epem{\mbox{e}^+\mbox{e}^- } 
\def\qqbar{\rm q \overline{q}}

\newlength{\capindent}
\newlength{\capwidth}
\newlength{\figwidth}
\newcommand{\icaption}[2][!*!,!]{\hspace*{\capindent}%
  \begin{minipage}{\capwidth}
    \ifthenelse{\equal{#1}{!*!,!}}%
      {\caption{#2}}%
      {\caption[#1]{#2}}
  \end{minipage}}

\begin{document}
\begin{titlepage}
\title{ Inclusive Lambda Production \\
        in Two-Photon Collisions at LEP}
\author{The L3 Collaboration}
\begin{abstract}
The reactions $\epem \ra \epem \Lambda \rm X$ and $\epem \ra 
\epem \overline{\Lambda} \rm X$ are studied using data collected at LEP with 
the L3 detector at centre-of-mass energies between $189$ and $209 \GeV$.
Inclusive differential cross sections are measured as a function of the
lambda transverse momentum, $p_t$, and pseudo-rapidity, $\eta$, in the ranges 
$0.4 \GeV < p_t< 2.5 \GeV$ and $|\eta| < 1.2$. The data are compared to Monte Carlo 
predictions. The differential cross section as a function of $p_t$ is well described 
by an exponential of the form $A \exp (-\pt/\langle\pt\rangle)$.
\end{abstract}
\submitted
\end{titlepage}

\section{Introduction}

Two-photon collisions are the main source of hadron production in high-energy $\rm \epem$ 
interactions at LEP via the process ${\rm e}^{+} {\rm e}^{-} \rightarrow {\rm e}^{+} 
{\rm e}^{-} \gamma ^{*} \gamma ^{*}  \rightarrow  {\rm e}^{+} {\rm e}^{-}  hadrons$. 
The outgoing electron and positron carry almost the full beam energy and their 
transverse momenta are usually so small that they escape, undetected, along the beam pipe. 
In this process, the negative four-momentum 
squared of the photons, $Q^2$, has an average value $\langle Q^2 \rangle 
\simeq 0.2 \GeV^2$ and they can therefore be considered 
as quasi-real. In the Vector Dominance Model (VDM), each photon can transform into a 
vector meson with the same quantum numbers, thus initiating a strong interaction 
process with characteristics similar to hadron-hadron interactions. This process 
dominates in the ``soft'' interaction region, where hadrons are produced with a 
low transverse momentum, $p_t$. Hadrons with high $p_t$ may also be produced 
by the QED process $\gamma \gamma \ra \qqbar$ (direct process) or by QCD processes 
originating from the partonic content of the photon (resolved processes). 

The L3 Collaboration has previously studied inclusive $\pi^0$, $\kos$ \cite{pz} 
and charged hadron \cite{ch} production in quasi-real two-photon collisions. 
In this Letter, the inclusive $\Lambda$ and $\overline{\Lambda}$ production\footnote{ 
If not mentioned otherwise, the symbol $\Lambda$ refers to both $\Lambda$ and 
$\overline{\Lambda}$, as both charge conjugate final state are analysed.} 
from quasi-real photons is studied for a centre-of-mass 
energy  of the two interacting photons, $\wgg$, greater than $5 \GeV$. 
The results are expressed in bins of transverse momentum, $p_t$, and
pseudo-rapidity, $\eta$, in the ranges $0.4 \GeV < p_t < 2.5 \GeV$ 
and $|\eta| < 1.2$. The $\eta$ range is further divided in two subsamples with 
$0.4 \GeV < p_t \leq 1 \GeV$ and $1 \GeV < p_t < 2.5 \GeV$. The data sample corresponds 
to a total integrated luminosity of 610 pb$^{-1}$ collected with the L3 detector 
\cite{det}, at $\rm \epem$ centre-of-mass energies $\sqrt{s}=189 - 209 \GeV$, with 
a luminosity weighted average value of $\sqrt{s}=198 \GeV$ . Results on inclusive 
$\Lambda$ production for a smaller data sample at lower $\sqrt{s}$ were 
previously reported by the TOPAZ Collaboration in the range 
$0.75 \GeV < p_t < 2.75 \GeV$ \cite{topaz}. The H1 Collaboration
investigated the $\Lambda$ photoproduction at HERA~\cite{h1}.

\section{Monte Carlo simulation}

The process $\epem \ra \epem hadrons$ is modelled with the PYTHIA
\cite{pythia} and PHOJET \cite{phojet} event generators with two times
more luminosity than the data. Both generators include the VDM, direct
and resolved processes.  In PYTHIA, including these processes for each
photon leads to six classes of interactions. A smooth transition
between these classes is then obtained by introducing a transverse
momentum scale. The two-photon luminosity function is based on the EPA
approximation~\cite{budnev} with a cut at the mass of the rho meson. The
SaS-1D parametrisation is used for the photon structure~\cite{sas1d}.
PHOJET relies on the Dual Parton Model~\cite{dpm}, with a soft and a
hard component. The two-photon luminosity functions is calculated in
the formalism of Reference~\citen{budnev}. The leading-order GRV
parametrisation is used for the photon structure~\cite{grv}.  For both
programs, matrix elements are calculated at the leading order and
higher-order terms are simulated by parton shower in the leading-log
approximation. The fragmentation is performed using the Lund string
fragmentation scheme as implemented in JETSET \cite{pythia}, which is also used
to simulate the hadronisation process. The strangeness suppression
factor in JETSET is chosen as 0.3, while a value $\alpha_S(m_{\rm
Z})=0.12$ is used for the strong coupling constrant.

From a study of Monte Carlo events, the hard component is found to be
larger than the soft component. Their ratio goes from around two at low values
of $p_t$ to around three at high values of $p_t$.

The following Monte Carlo generators are used to simulate the  
background processes: KK2f\cite{KK2f} for the annihilation process $\epem \rightarrow \rm \qqbar \,
(\gamma $); KORALZ \cite{KORALZ} for $\epem \rightarrow \tau^{+} \tau^{-}(\gamma )$; 
KORALW \cite{KORALW} for $\epem \rightarrow \rm{W}^{+} \rm{W}^{-}$  and  
DIAG36 \cite{DIAG36} for $\epem \ra  \epem \tau^{+} \tau^{-}$. 
The response of the L3 detector is simulated using the GEANT \cite{GEANT} and GHEISHA \cite{GEISHA} 
programs. Time dependent detector inefficiencies, as monitored during each data taking period, 
are included in the simulations. All simulated events are passed through the same reconstruction 
program as the data.

\section{Event selection}

\begin{figure}
\begin{center}
\epsfig{file=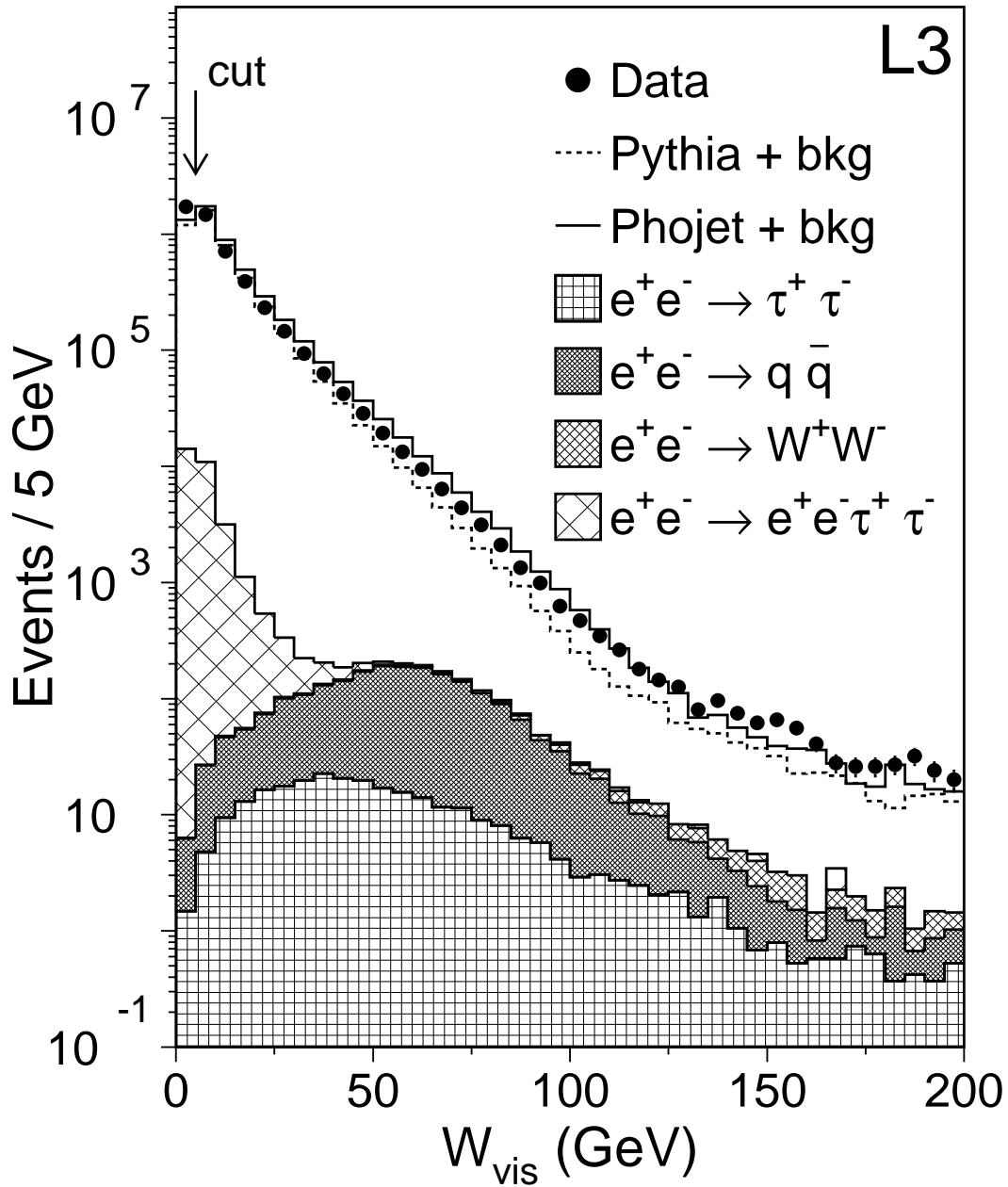,width=14cm} 
\end{center}
\caption {The distribution of the visible mass $W_{vis}$. 
The Monte Carlo distributions are normalized to the data 
luminosity. Various contributions to the background (bkg) are shown as cumulative
histograms.}
\label{hadsel}
\end{figure}

Two-photon interaction events are mainly collected by the track triggers 
\cite{trigtrack}, with a low transverse momentum threshold of about 150 \MeV, 
and the calorimetric energy trigger \cite{trigener}. The selection of  
${\rm e}^{+} {\rm e}^{-} \rightarrow {\rm e}^{+} {\rm e}^{-} hadrons$ 
events \cite{l3tot} is based on information from the central tracking detectors and 
the electromagnetic and hadronic calorimeters. It consists of:

\begin{itemize}

\item A multiplicity cut. To select hadronic final states, at  
least six objects must be detected, where an object can be a 
track satisfying minimal quality requirements or an isolated calorimetric cluster of 
energy greater than 100 \MeV.

\item 
Energy cuts. The total energy deposited in the calorimeters
must be less than 40\% of $\sqrt{s}$ to suppress events from the 
$\epem \ra \rm q \overline{q} (\gamma)$ and $\epem \ra \tau^+ \tau^- (\gamma)$ processes.
In addition, the total energy in the electromagnetic calorimeter is required to be greater 
than $500 \MeV$ in order to suppress beam-gas and beam-wall interactions
and less than $50 \GeV$ to remove events from the annihilation process 
$\epem \ra \rm q \overline{q} (\gamma)$.

\item
An anti-tag condition. Events with a cluster in the luminosity monitor, which 
covers the angular region 31 mrad $<\theta<$ 62 mrad, with an electromagnetic shower 
shape and energy greater than $30 \GeV$ are excluded from the analysis. In addition, events with an electron
scattered above 62 mrad are rejected by the energy cuts.

\item
A mass cut. The invariant mass of all visible particles, $W_{vis}$, must 
be greater than $5 \GeV$. In this calculation, the pion mass is attributed to 
tracks while isolated electromagnetic clusters are treated as massless. The distribution 
of $W_{vis}$ for data and Monte Carlo after all other cuts are applied is shown in Figure 
\ref{hadsel}. Values of $W_{vis}$ up to $200 \GeV$ are accessible.

\end{itemize}

About 3 million hadronic events are selected by these criteria with an 
overall efficiency of 45\%. The background level of this sample is 
less than 1\% and is mainly due to the $\epem \ra \rm q 
\overline{q}(\gamma)$ and $\epem \ra \epem \tau^+ \tau^-$ processes.
The background from beam-gas and beam-wall interactions is found to be negligible.

The $\Lambda$ baryons are reconstructed using the decay \mbox{$\Lambda \ra {\rm p} 
\pi$}. Secondary decay vertices are selected which are formed by two oppositely charged tracks. 
The secondary vertices must satisfy the following criteria:

\begin{itemize}

 \item  The distance, in the plane transverse to the beam direction, 
 between the secondary vertex and the primary $\epem$ interaction point 
 must be greater than 3 mm. 

 \item The angle between the total transverse momentum vector of 
 the two outgoing tracks and the direction in the transverse plane between the primary interaction 
 point and the secondary vertex must be less than 100 mrad. 
\end{itemize}

\begin{figure}
\begin{center}
\epsfig{file=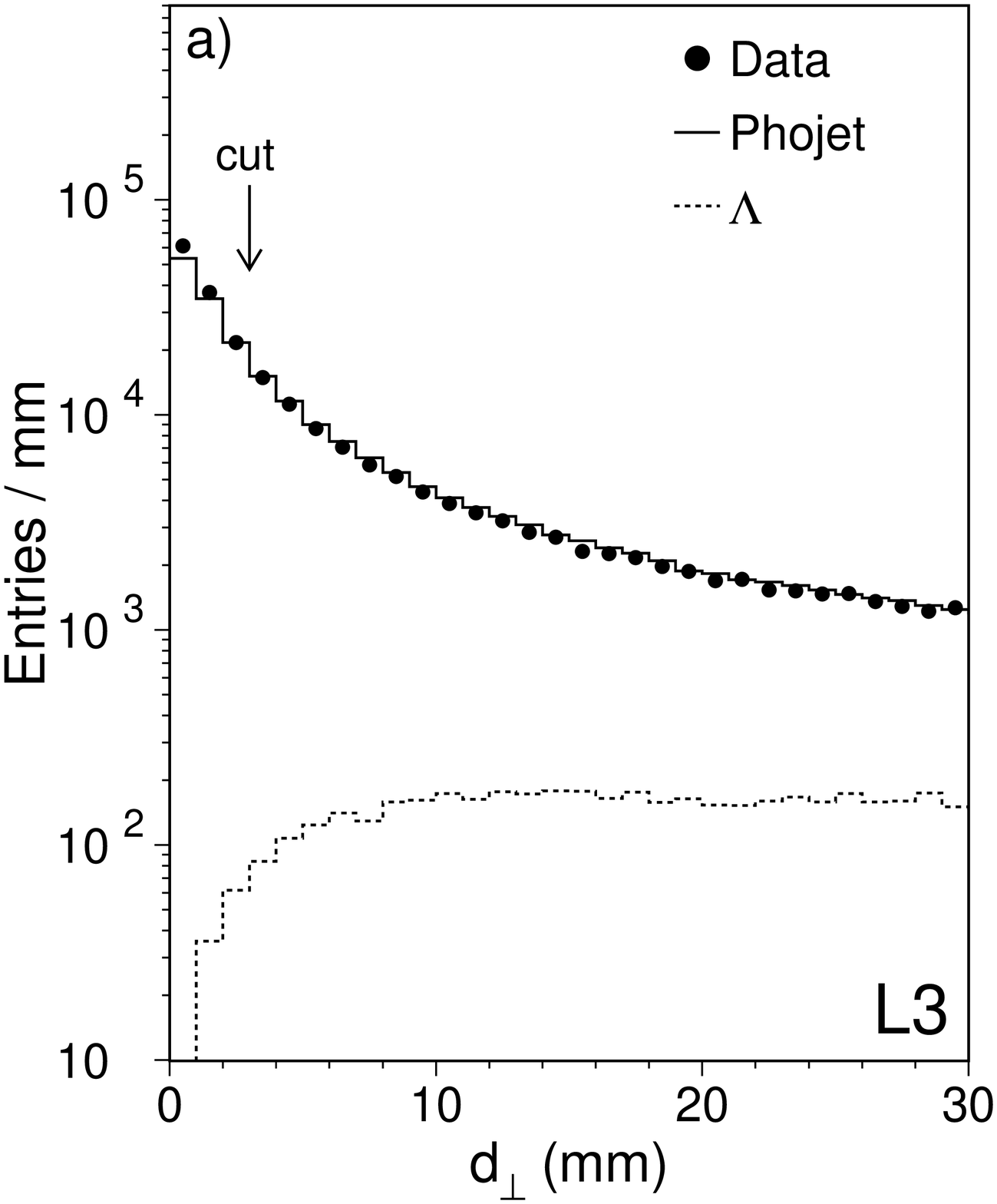,width=7.5cm} 
\epsfig{file=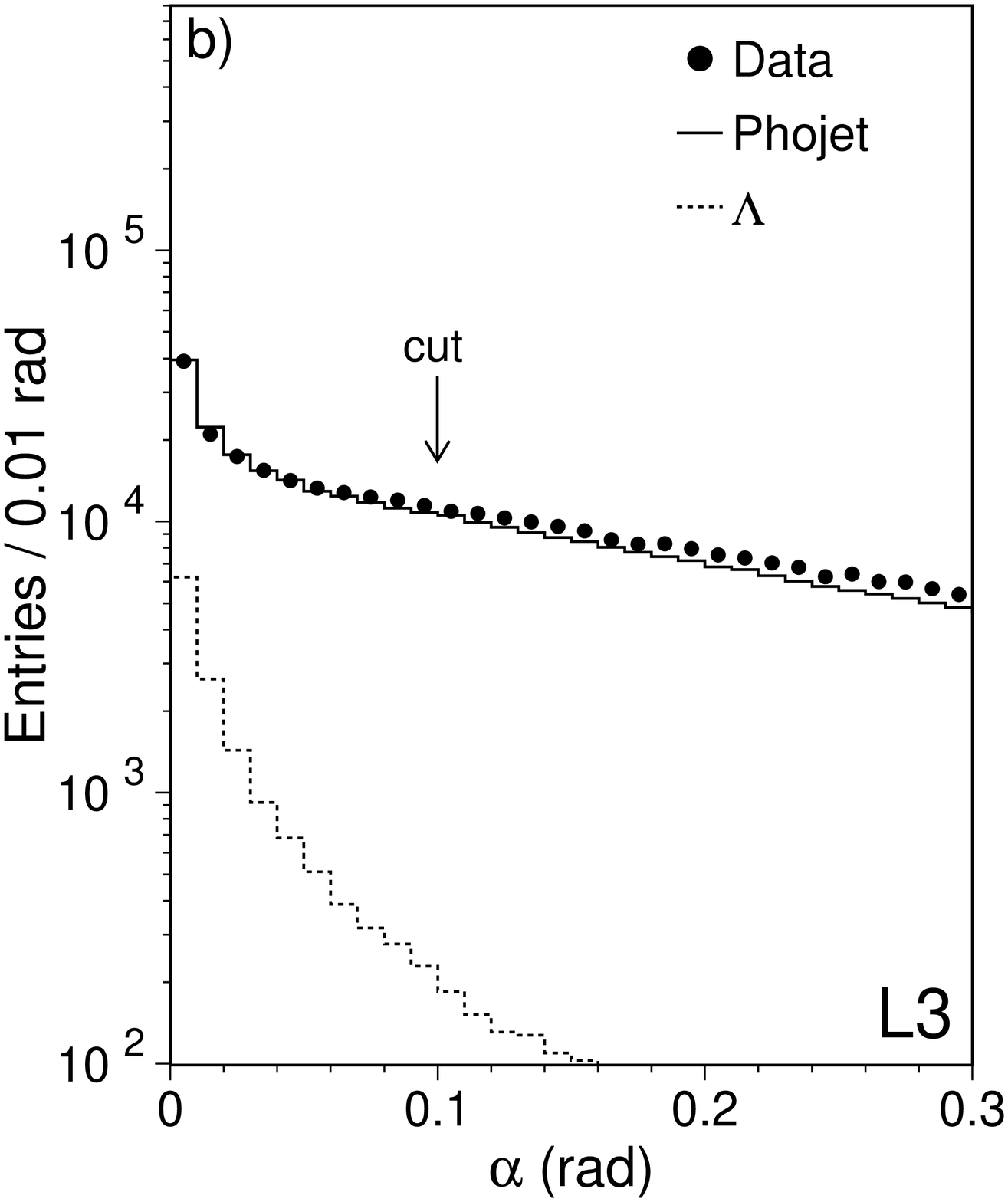,width=7.5cm} 
\end{center}
\caption {Distribution of the variables used for the selection of secondary vertices: a) the distance, 
in the plane transverse to the beam direction, between the secondary vertex and the primary 
$\epem$ interaction point, $\rm d_\perp$, and b) the angle between the total transverse momentum vector of 
 the two outgoing tracks and the direction in the transverse plane between the primary interaction 
 point and the secondary vertex, $\alpha$. In each plot, all other selection criteria are applied. The predictions 
of the PHOJET Monte Carlo are shown as the full line and the contribution due to $\Lambda$ baryons as the
dashed line. The Monte Carlo distributions are normalized to the data luminosity.}
\label{cutvtx}
\end{figure}

The distributions of these variables are presented in Figure \ref{cutvtx}.
The proton is identified as the track with the largest momentum. Monte Carlo studies show that this association is correct for 
more than 99\% of the vertices. In addition, the d$E$/d$x$ measurement of both proton and 
pion candidates must be consistent with this hypothesis with a confidence level 
greater than 0.01.

\begin{figure}
\begin{center}
\epsfig{file=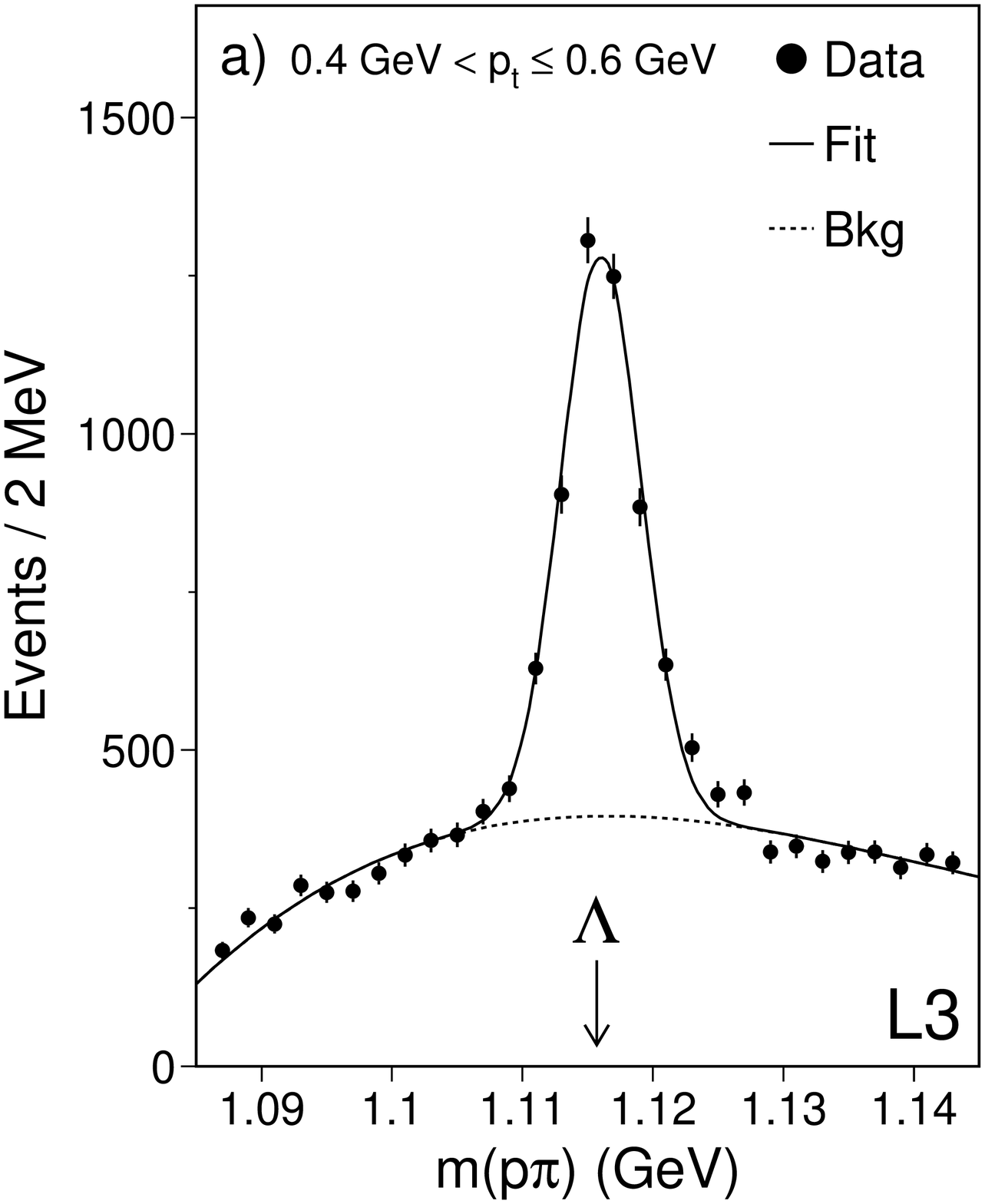,width=7.5cm} \hspace{0.5cm}
\epsfig{file=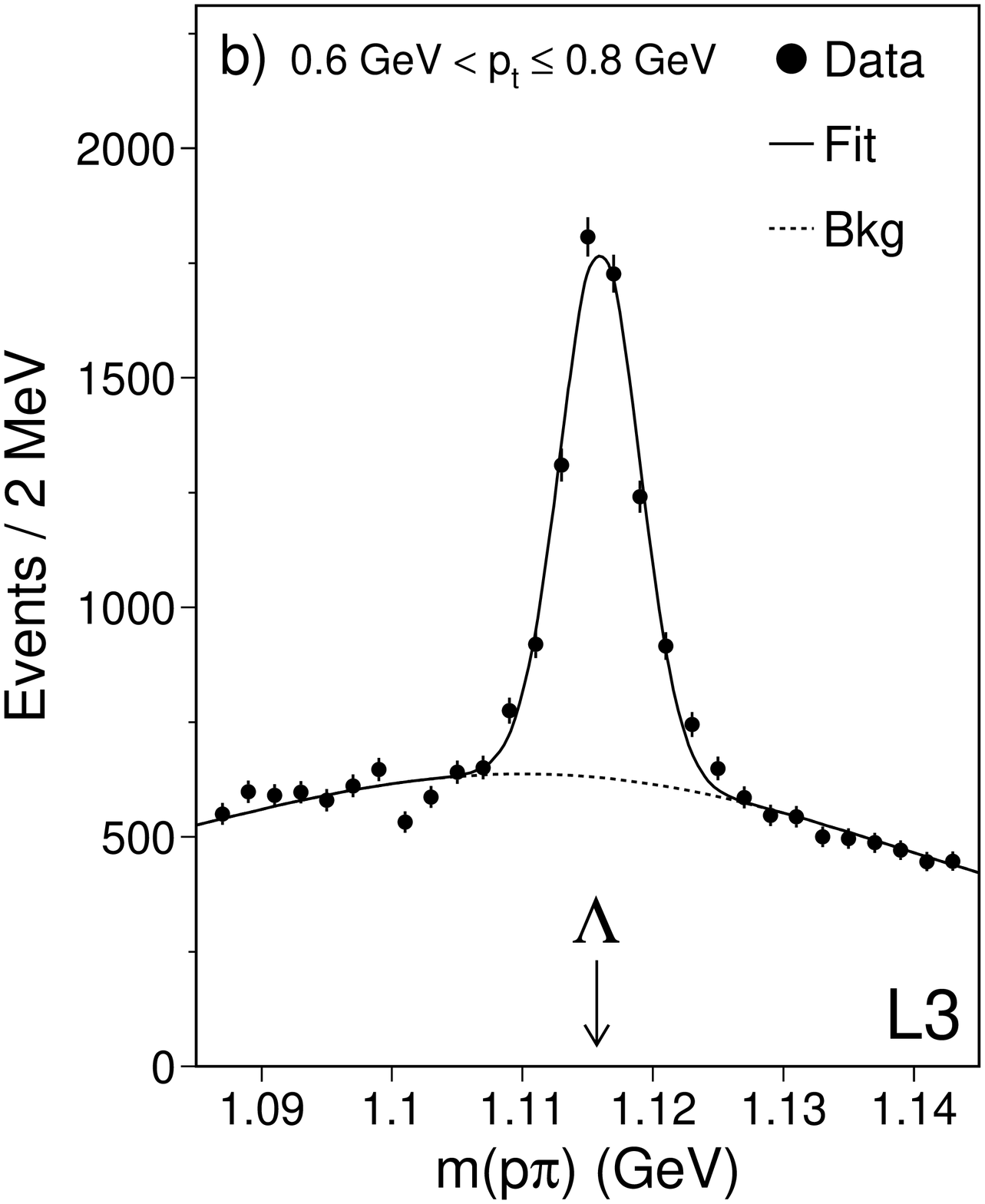,width=7.5cm} \vspace{0.5cm}\\
\epsfig{file=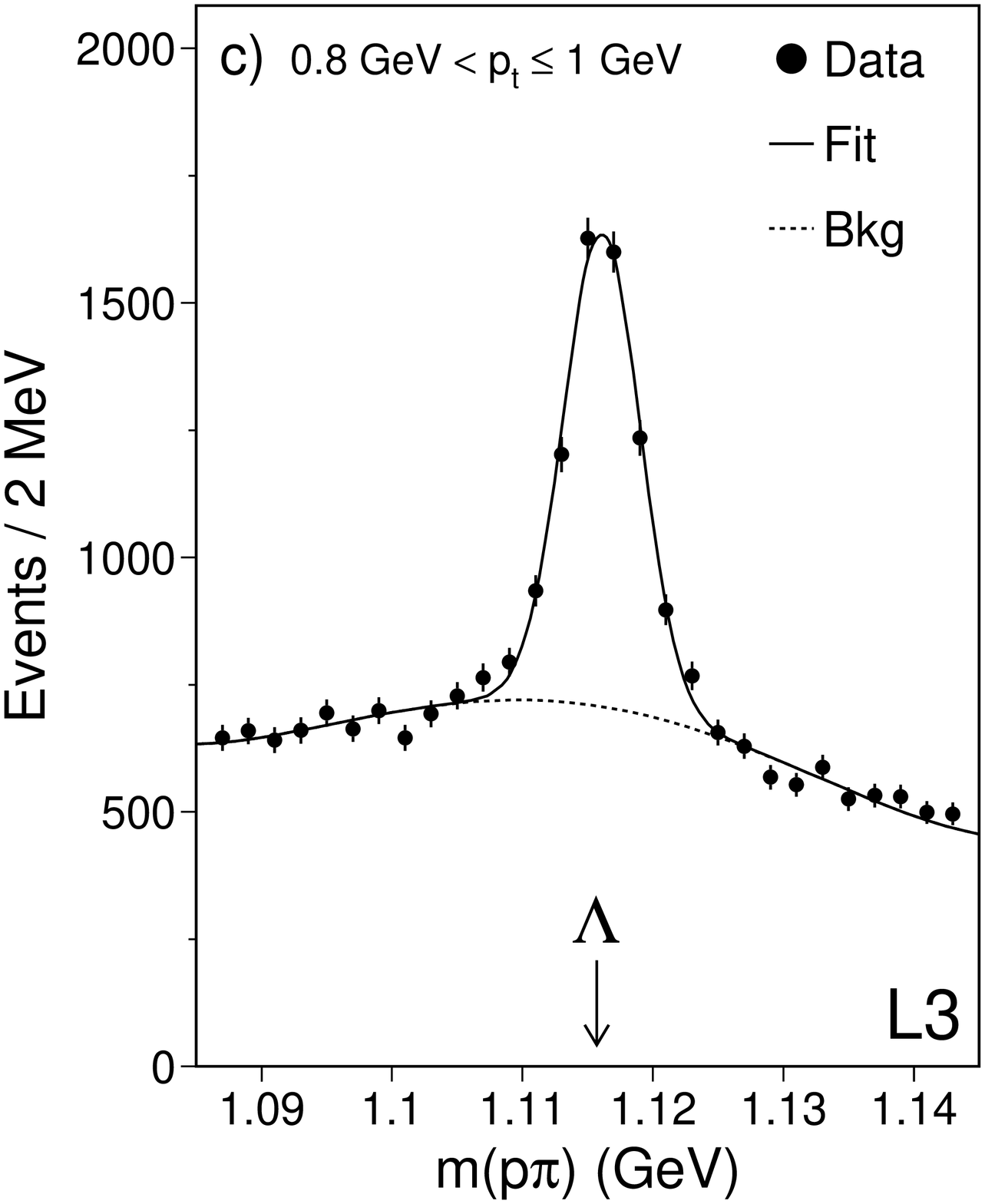,width=7.5cm} \hspace{0.5cm}
\epsfig{file=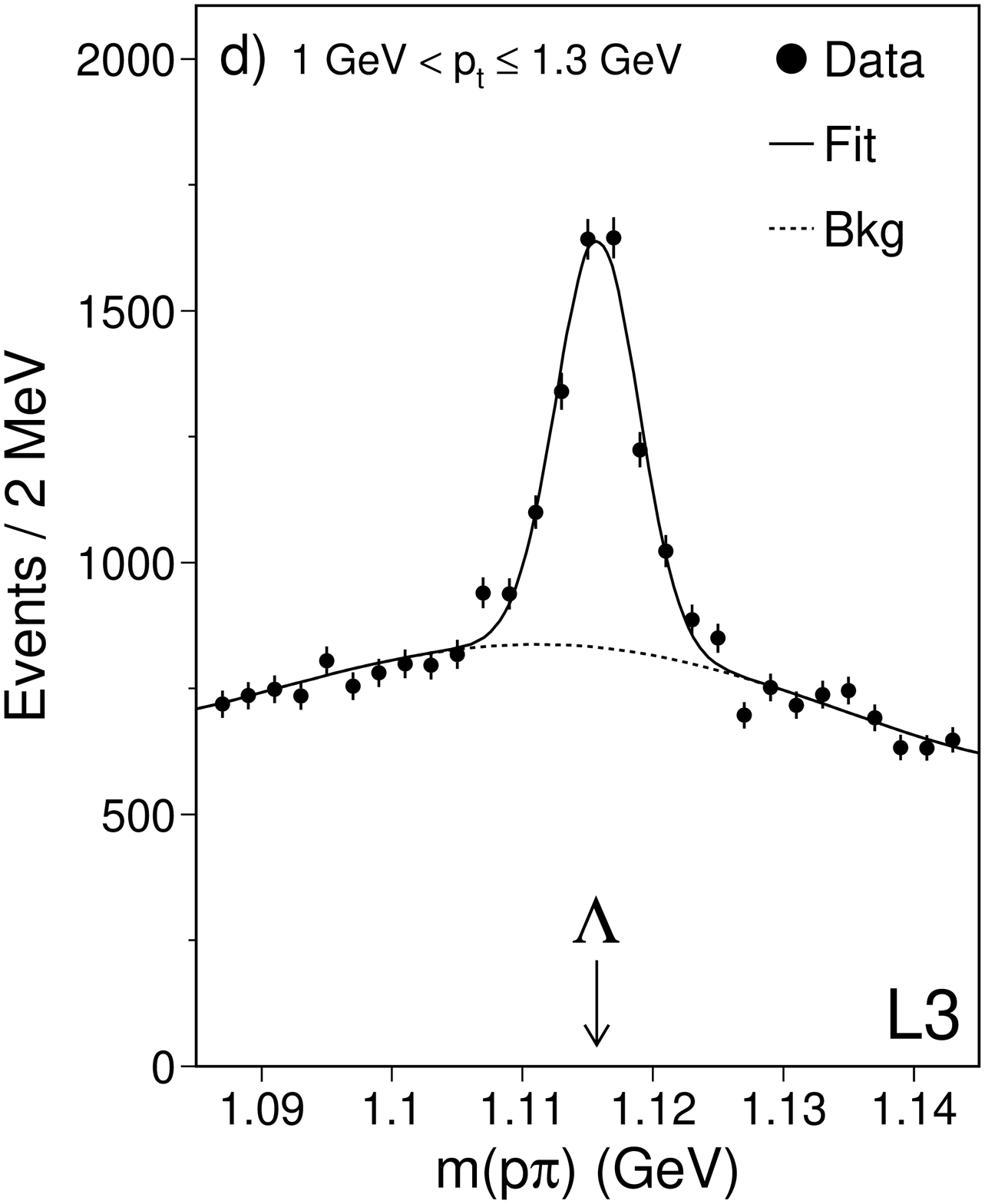,width=7.5cm} 
\end{center}
\caption {The invariant mass of the $\rm p \pi$ system for a) $0.4 \GeV < p_t \leq 0.6 \GeV $, b) 
$0.6 \GeV < p_t \leq 0.8 \GeV $, c)  $0.8 \GeV < p_t \leq 1 \GeV $ and d) $1 \GeV  < p_t \leq 1.3 \GeV$ . 
The signal is modelled with a Gaussian and the background by a fourth-degree Chebyshev 
polynomial.}
\label{lambdapl}
\end{figure}

\begin{figure}
\begin{center}
\epsfig{file=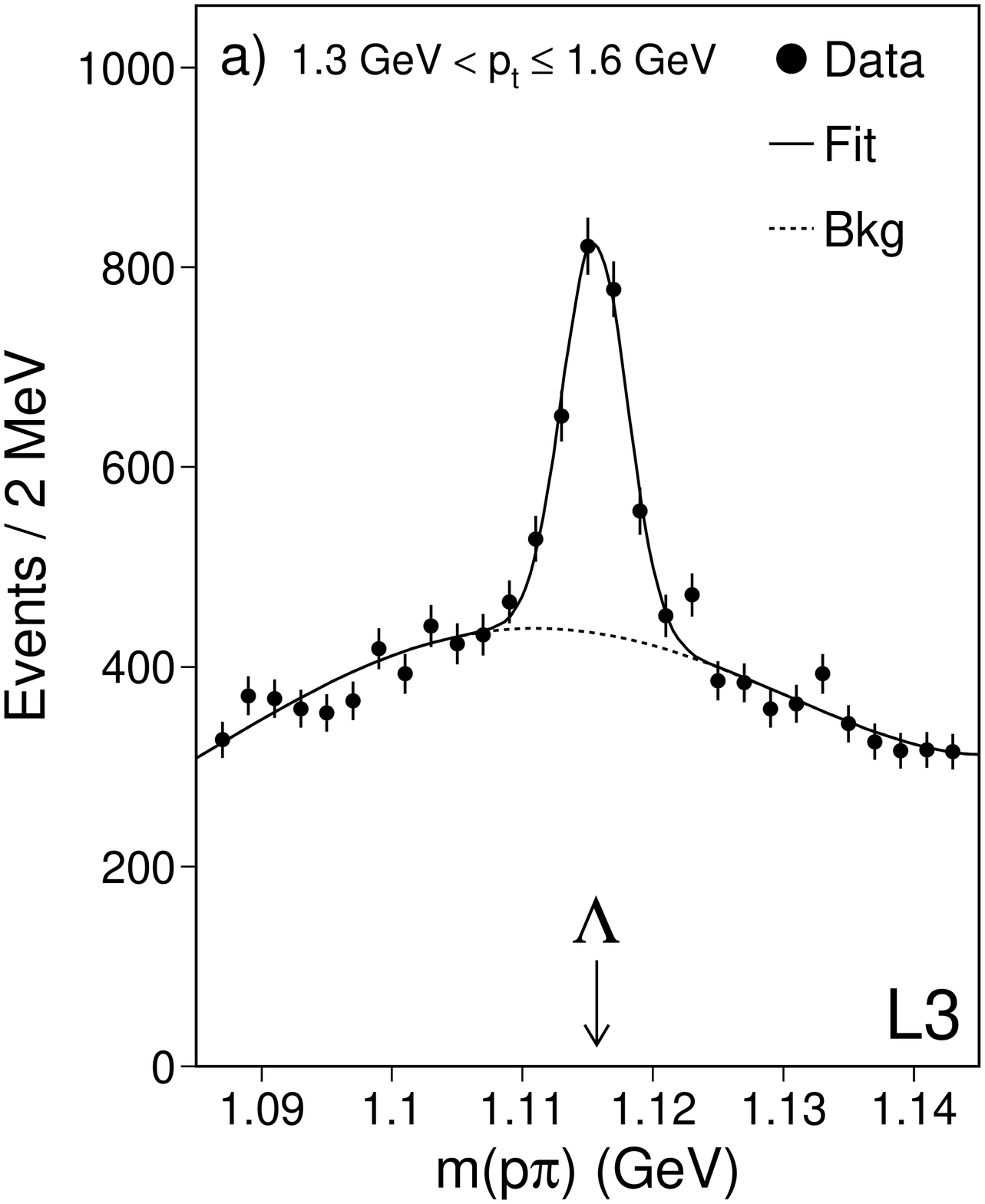,width=7.5cm} \hspace{0.5cm}
\epsfig{file=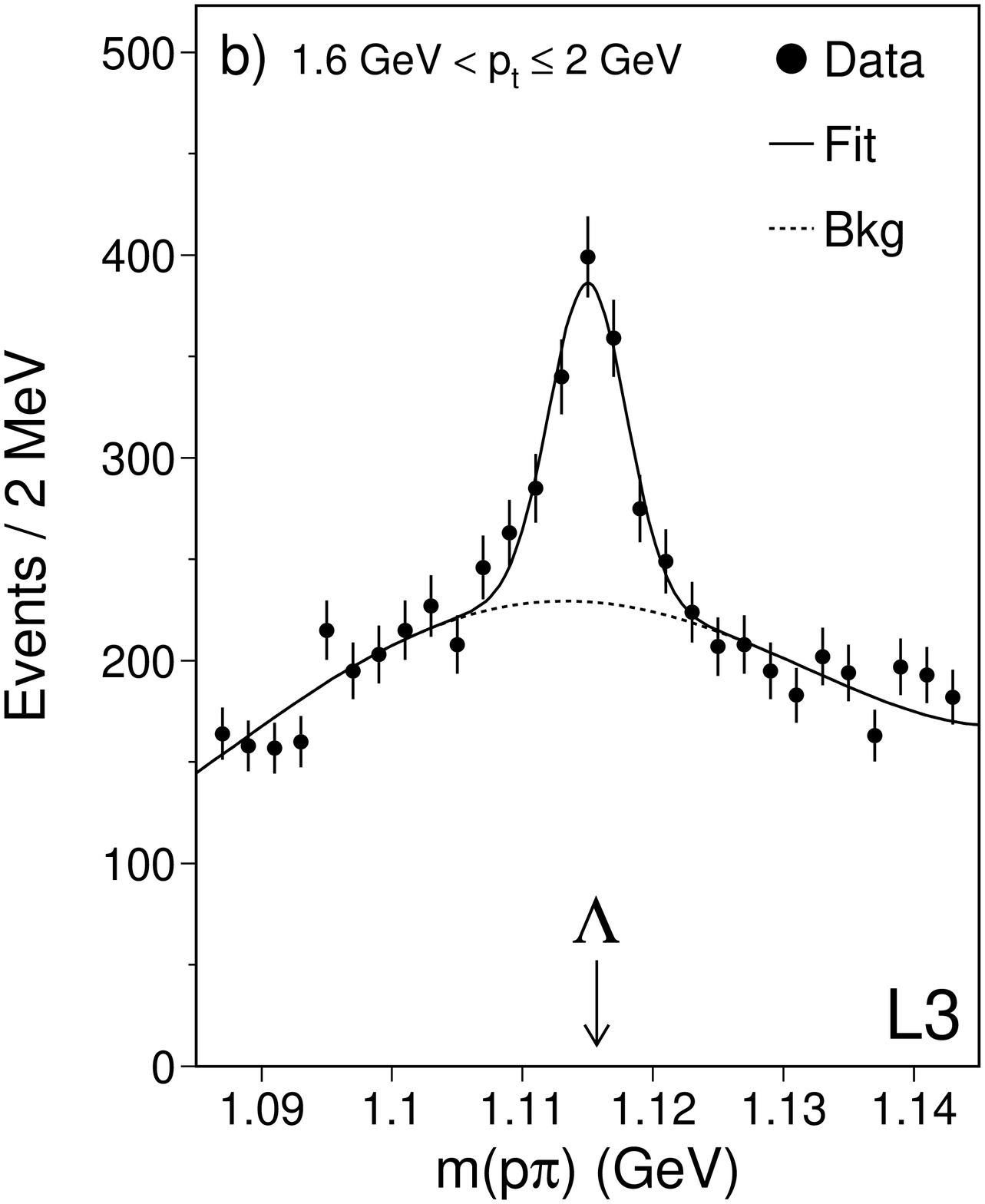,width=7.5cm} \vspace{0.5cm}\\
\epsfig{file=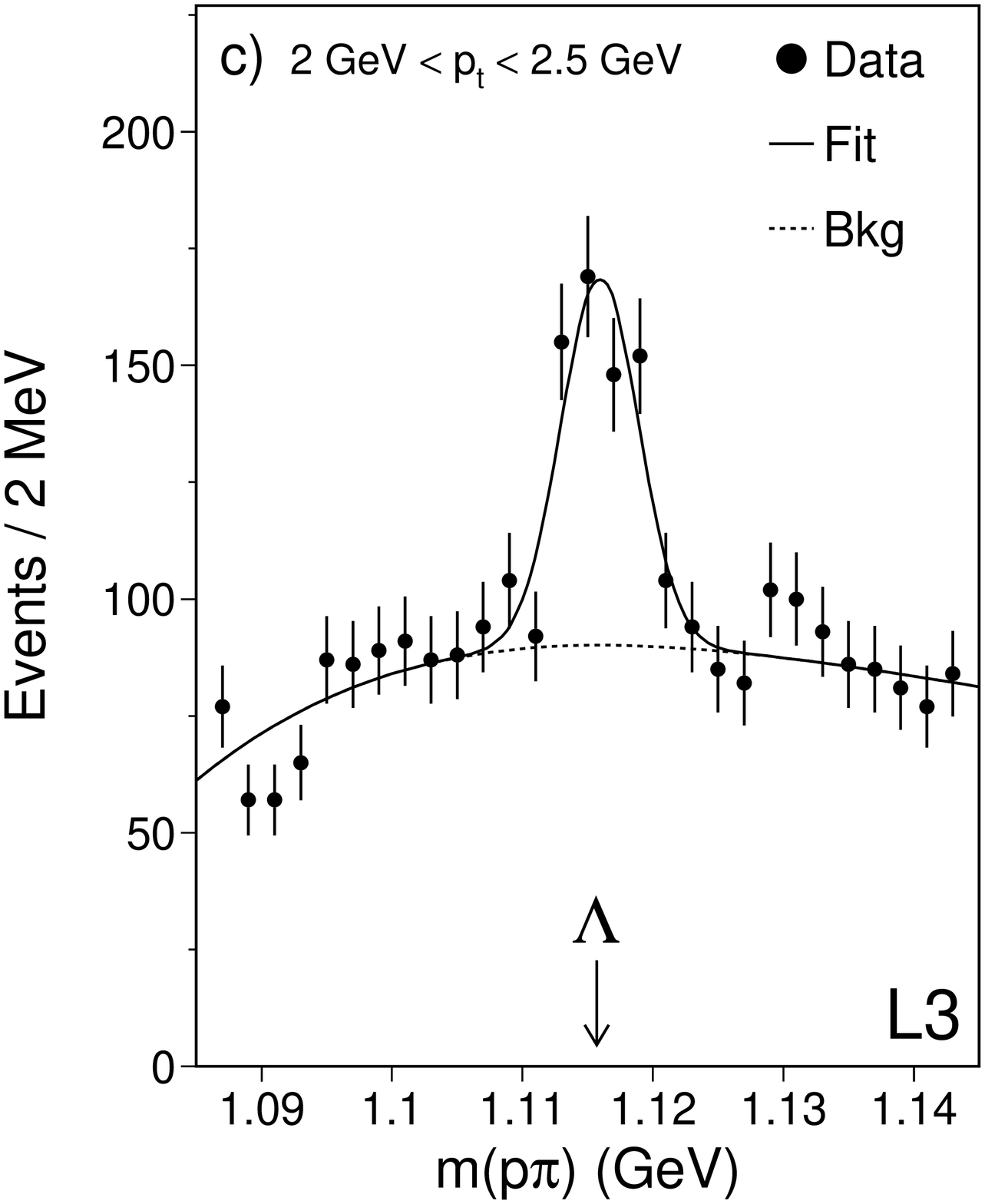,width=7.5cm} \hspace{0.5cm}
\end{center}
\caption {The invariant mass of the $\rm p \pi$ system for a) $1.3 \GeV < p_t \leq 1.6 \GeV$, b) 
$1.6 \GeV  < p_t \leq 2 \GeV $ and c) $2 \GeV  < p_t < 2.5 \GeV $. 
The signal is modelled with a Gaussian and the background by a fourth order Chebyshev polynomial.}
\label{lambdapl2}
\end{figure}

After these cuts, about 70000 events are selected. The distribution 
of the invariant mass of the ${\rm p} \pi$ system, $m({\rm p} \pi)$, 
shows a clear $\Lambda$ peak, as shown in Figures \ref{lambdapl} and 
\ref{lambdapl2} for the different $p_t$ bins listed in Table \ref{tab1}.
The resolution of $m({\rm p} \pi)$ is found to be around $3 \MeV$ and 
is well reproduced by Monte Carlo simulations.

\section {Differential cross sections}

The differential cross sections for $\Lambda$ production as a function of $p_t$ and $|\eta|$ are measured 
for $\wgg > 5 \GeV$, with a mean value of $30 \GeV$, and a photon virtuality 
$Q^2 < 8 \GeV^2$ with $\langle Q^2 \rangle \simeq 0.2 \GeV^2$. This phase space 
is defined by cuts at the generator level of the Monte Carlo.

The number of $\Lambda$ baryons in each $p_t$ and $|\eta|$ bin is evaluated by means 
of a fit to the $ m({\rm p} \pi)$ spectrum in the interval $1.085 \GeV 
< m({\rm p} \pi) < 1.145 \GeV$. The signal is modelled with a Gaussian and the 
background by a fourth-degree Chebyshev polynomial. All parameters, 
including the mass and width of the peak, are left free. The results 
are given in Tables \ref{tab1}, \ref{tab2} and \ref{tab3}.

The overall efficiencies, also listed in Tables \ref{tab1}, \ref{tab2} and \ref{tab3}, include
reconstruction and trigger efficiencies and take into account the 64\% branching fraction of the 
decay $\Lambda \ra {\rm p} \pi$. The reconstruction efficiency, which includes effects of the 
acceptance and the selection cuts, is calculated with the PHOJET and PYTHIA Monte Carlo 
generators. As both generators reproduce well the shapes of the experimental 
distributions of hadronic two-photon production \cite{l3tot}, their average is used
to calculate the selection efficiency. The efficiency does not depend on the $Q^2$ cut-off.   
The track trigger efficiency is calculated for each data taking period 
by comparing the number of events accepted by the track 
and the calorimetric energy triggers. The efficiency of the higher level triggers 
is measured using prescaled events. The total trigger efficiency increases 
from 82\% for $p_t<0.4 \GeV$ to 85\% in the high $p_t$ region.

The differential cross section as a function of $p_t$ is first measured for the 
three different data samples collected in 1998, 1999 and 2000 at different values of $\sqrt{s}$ and 
for different trigger and machine background conditions. Good agreement 
is obtained between the three measurements. The different data samples 
are therefore combined into a single data sample.

Sources of systematic uncertainties on the cross section measurements are:
background subtraction, scale and resolution uncertainties, Monte Carlo modelling and 
the accuracy of the trigger efficiency measurement. Their contributions are 
listed in Table \ref{tabsyst}. The dominant source of systematic uncertainty 
is due to background subtraction. It is estimated using different background 
parameterizations and fit intervals in the fitting procedure. The
scale and resolution uncertainties are assessed by varying the
selection cuts. The main contributions arises from the secondary vertex selection (3.2\%) 
and the proton and pion identification criteria (2.5\%). The uncertainty 
due to the selection of $\epem \ra \epem hadrons $ events is 1\%. 
The Monte Carlo modelling uncertainty, taken as half the relative difference between 
PHOJET and PYTHIA, increases with $p_t$ from 1\% to 5\%. A systematic uncertainty of 2\% 
is assigned to the determination of the trigger efficiency, which takes into account the determination procedure 
and time stability.

\section{Results}

The average multiplicity of $\Lambda$ baryons in the range $\wgg > 5
\GeV$, $0.4 \GeV  < p_t <2.5 \GeV$ and $|\eta|<$ 1.2 
is $(1.57 \pm 0.11)\times 10^{-2}$ per $\epem \ra \epem hadrons$ event. The uncertainty 
includes both statistical and systematic uncertainties. This result is below the 
value $(1.80 \pm 0.01)\times 10^{-2}$ predicted by PHOJET and above 
the value $ (1.43 \pm 0.01)\times 10^{-2}$ predicted by PYTHIA. The ratio 
of $\Lambda$ to $\overline{\Lambda}$ baryons, as determined from the charge of the most energetic 
track of the vertex, is found to be $\rm N(\overline{\Lambda}) / N(\Lambda) = 0.99 \pm
0.04$. 

\begin{figure}
\begin{center}
\epsfig{file=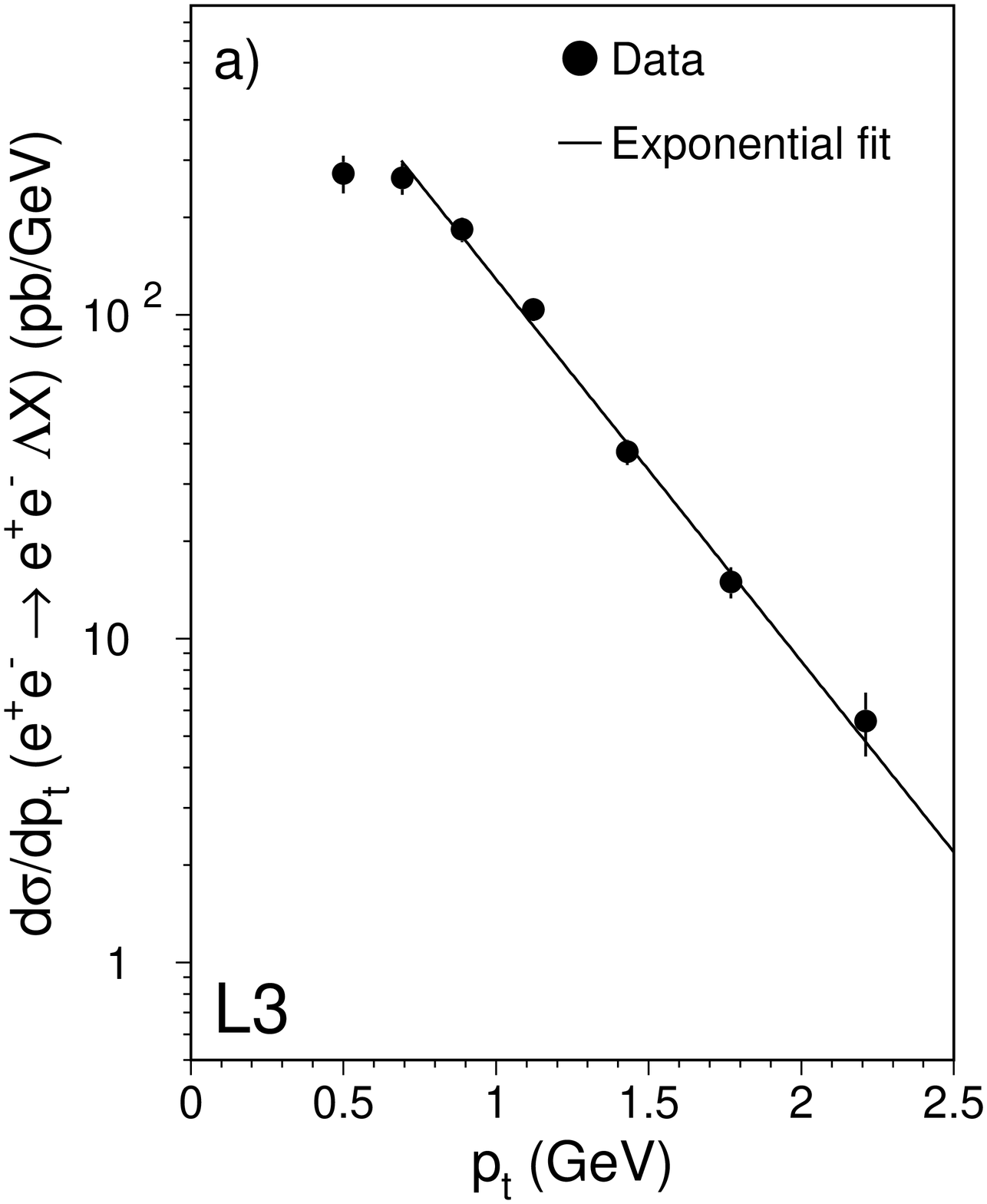,width=7.5cm} \hspace{0.5cm}
\epsfig{file=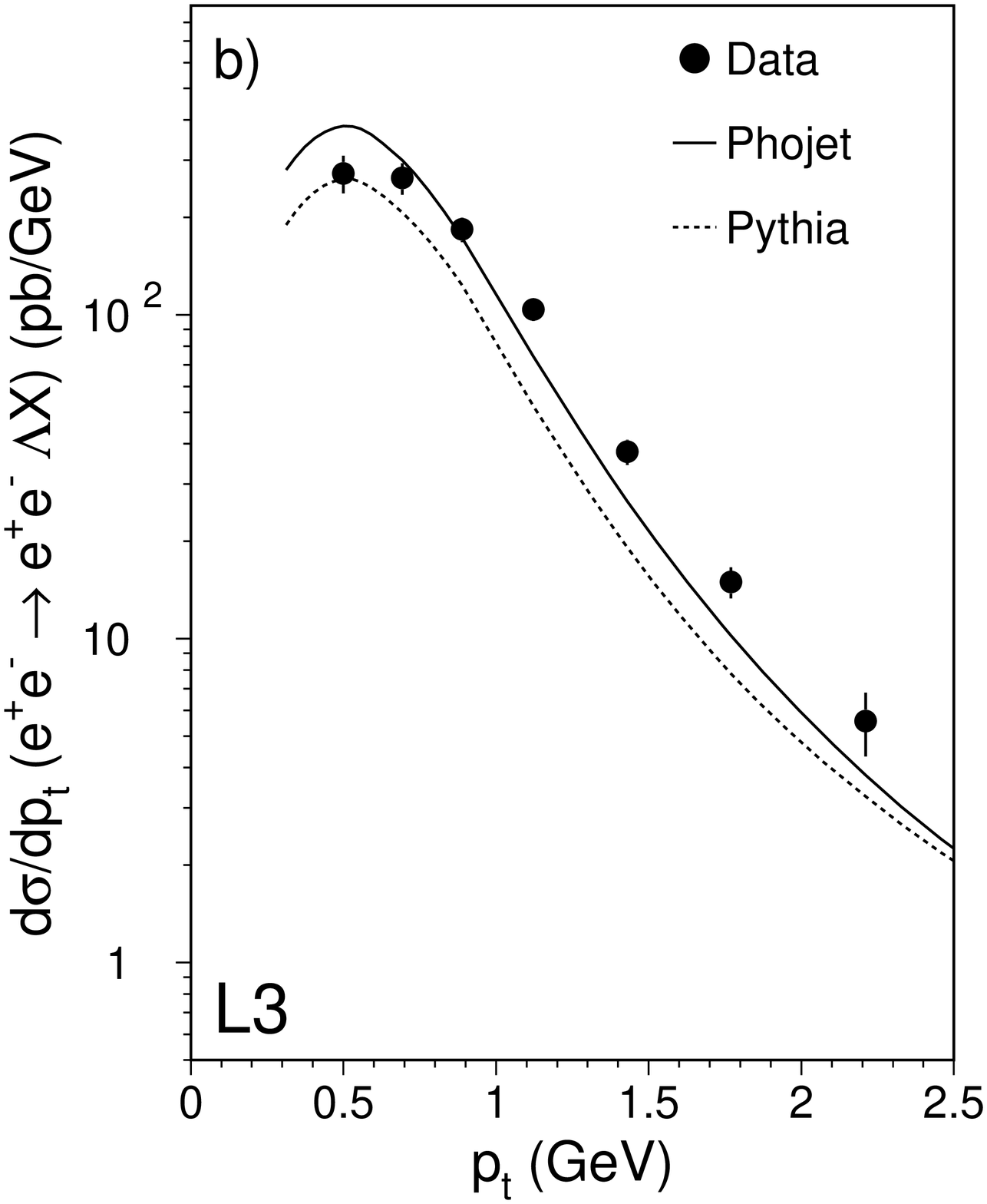,width=7.5cm} 
\end{center}
\caption {The differential cross section as a function of $p_t$ for the 
$\epem \ra \epem \Lambda  \rm X $ and $\epem \ra \epem \overline{\Lambda} \rm X $ 
processes for $|\eta|<1.2$: a) with the exponential fit described in the text and 
b) compared to the predictions of the PHOJET and PYTHIA Monte Carlo programs.}
\label{crosspt}
\end{figure}

The sum of the differential cross sections for the $\epem \ra \epem
\Lambda \rm X $ and $\epem \ra \epem \overline{\Lambda} \rm X $
processes as a function of $p_t$ for $|\eta| < 1.2$ is presented in
Table \ref{tab1} and Figure \ref{crosspt}. Mass effects explain the
lower value obtained in the first bin. The behaviour of the cross
section for $0.75 \GeV < p_t < 2.5 \GeV$ is well described by an
exponential of the form $A \exp (-\pt/\langle\pt\rangle)$, as shown in
Figure \ref{crosspt}a, with a mean value \mbox{$\langle\pt\rangle =
368 \pm 17 \MeV$}. For comparison, the values \mbox{$\langle\pt\rangle
\simeq $ 230 \MeV} and \mbox{$\langle\pt\rangle \simeq 290 \MeV$} are
obtained for inclusive $\pi^0$ and $\rm K^0_S$ production,
respectively \cite{pz}.

The data are compared to 
the PHOJET and PYTHIA Monte Carlo predictions in Figure \ref{crosspt}b. The 
region $p_t<0.6 \GeV$ is well described by PYTHIA, whereas PHOJET 
overestimates the cross section. On the contrary, the region $0.6 \GeV  \leq p_t \leq  1 
\GeV $ is better reproduced by PHOJET while PYTHIA lies below the data. For
\mbox{$p_t>1$ \GeV } both PYTHIA and PHOJET underestimate the
data. This level of agreement is not unusual in two-photon physics.

\begin{figure}
\begin{center}
\epsfig{file=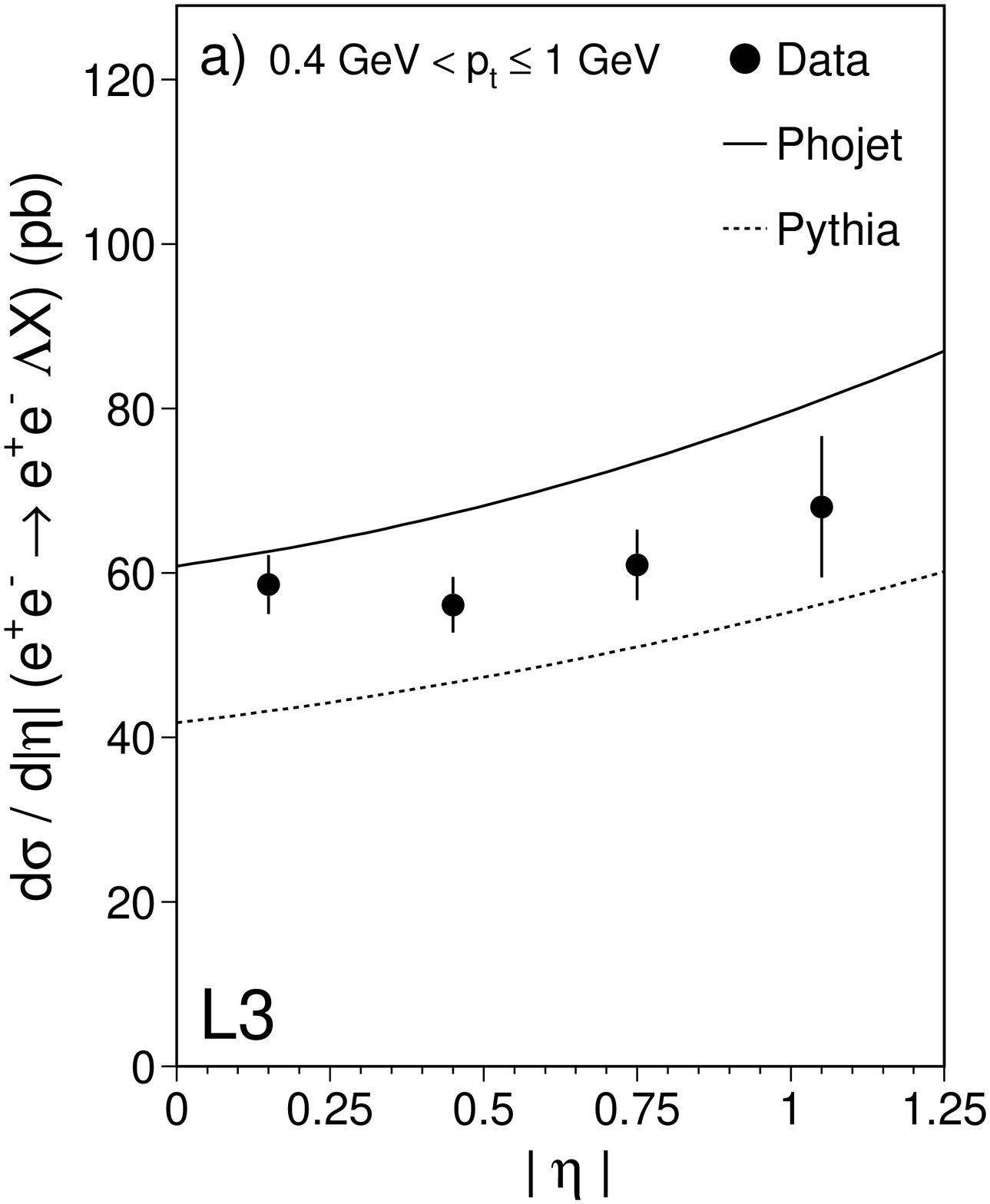,width=7.5cm} \hspace{0.5cm}
\epsfig{file=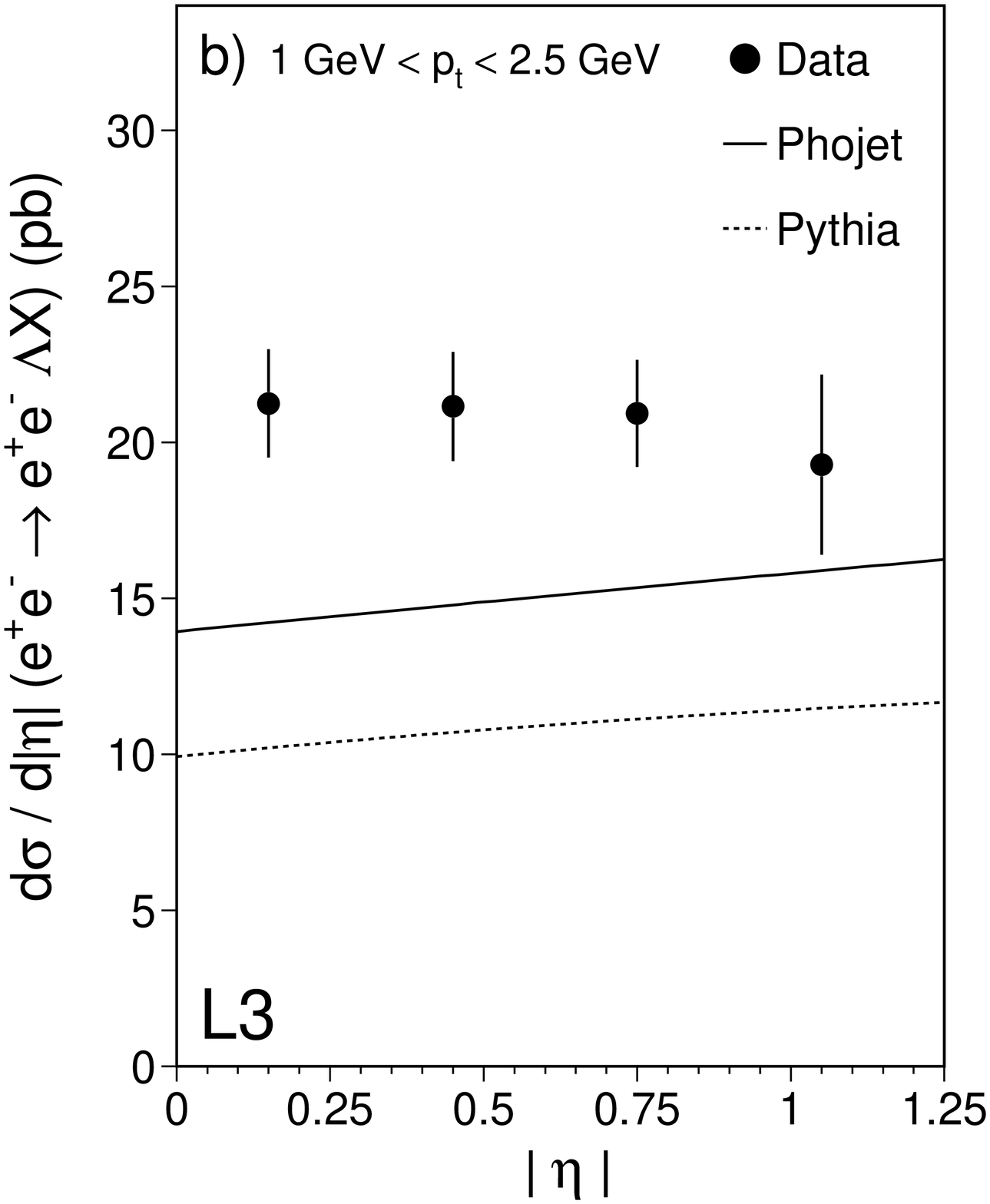,width=7.5cm} 
\end{center}
\caption {The differential cross section as a function of $|\eta|$ for the
$\epem \ra \epem \Lambda  \rm X $ and $\epem \ra \epem \overline{\Lambda} \rm X $ 
processes for: a) $0.4 \GeV  < p_t \leq 1 \GeV$ and b) $1 \GeV < p_t < 2.5 \GeV$. The data are compared to the predictions 
of the PHOJET and PYTHIA Monte Carlo programs.}
\label{crosseta}
\end{figure}

The differential cross sections as a function of $|\eta|$ for $0.4 \GeV  
<p_t \leq 1 \GeV $ and $1 \GeV < p_t < 2.5 \GeV $ are given in Tables \ref{tab2} 
and \ref{tab3} and shown in Figure \ref{crosseta}. Both Monte Carlo programs describe 
well the almost uniform $\eta$ shape, while the size of the discrepancy on the absolute normalization 
depends on the $p_t$ range.



\newpage
\bibliographystyle{l3style}
\bibliography{lambdax}

\begin{thebibliography}{10}

\bibitem{pz}
L3 Collab., P. Achard {\it et al.}, Phys. Lett. {\bf B 524} (2002) 44.

\bibitem{ch}
L3 Collab., P. Achard {\it et al.}, Phys. Lett. {\bf B 554} (2003) 105.

\bibitem{det}
L3 Collab., B. Adeva {\it et al.}, Nucl. Instr. Meth. {\bf A 289} (1990) 35;\\
  L3 Collab., O. Adriani {\it et al.}, Phys. Rept. {\bf 236} (1993) 1;\\ M.
  Chemarin {\it et al.}, Nucl. Instr. Meth. {\bf A 349} (1994) 345;\\ M.
  Acciarri {\it et al.}, Nucl. Instr. Meth. {\bf A 351} (1994) 300;\\ I. C.
  Brock {\it et al.}, Nucl. Instr. Meth. {\bf A 381} (1996) 236;\\ A. Adam {\it
  et al.}, Nucl. Instr. Meth. {\bf A 383} (1996) 342.

\bibitem{topaz}
TOPAZ Collab., R. Enomoto {\it et al.}, Phys. Lett. {\bf B 347} (1995) 179.

\bibitem{h1}
H1 Collab., C. Adloff {\it et al.}, Z. Phys. {\bf C 76} (1997) 213.

\bibitem{pythia}
{PYTHIA version 5.722 and JETSET version 7.409 are used with default options;\\
  T. Sj\"ostrand, Comp. Phys. Comm. {\bf 82} (1994) 74.}

\bibitem{phojet}
PHOJET version 1.05c is used with default options; \\ R. Engel, Z. Phys. {\bf C
  66} (1995) 203;\\ R. Engel and J. Ranft, \PR {\bf D 54} (1996) 4244.

\bibitem{budnev}
V.M. Budnev {\it et al.}, Phys. Rept. {\bf 15} (1974) 181.

\bibitem{sas1d}
G. Schuler and T. Sj{\"o}strand, Z. Phys. {\bf C 68} (1995) 607.

\bibitem{dpm}
A. Capella {\it et al.}, Phys. Rep. {\bf 236} (1994) 225.

\bibitem{grv}
M. Gluck, E. Reya and A. Vogt, Phys. Rev. {\bf D 45} (1992) 3986;\\ Phys.
  Rev.{\bf D 46} (1992) 1973.

\bibitem{KK2f}
KK2f version 4.12 is used;\\ S. Jadach, B. F. L. Ward and Z. W\c{a}s, Comp.
  Phys. Comm. {\bf 130} (2000) 260.

\bibitem{KORALZ}
KORALZ version 4.04 is used;\\ S. Jadach, B. F. L. Ward and Z. W\c{a}s, Comp.
  Phys. Comm. {\bf 79} (1994) 503.

\bibitem{KORALW}
KORALW version 1.33 is used;\\ M. Skrzypek {\it et al.}, Comp. Phys. Comm. {\bf
  94} (1996) 216.

\bibitem{DIAG36}
DIAG36 Monte Carlo;\\ F. A. Berends, P. H. Daverfeldt and R. Kleiss, \NP {\bf B
  253} (1985) 441.

\bibitem{GEANT}
GEANT version 3.15 is used;\\ R. Brun {\it et al.}, CERN report CERN DD/EE/84-1
  (1984), revised 1987.

\bibitem{GEISHA}
H. Fesefeldt, RWTH Aachen report PITHA 85/2 (1985).

\bibitem{trigtrack}
P. B\'en\'e {\it et al.}, Nucl. Inst. Meth. {\bf A 306} (1991) 150; \\ D. Haas
  {\it et al.}, Nucl. Inst. Meth. {\bf A 420} (1991) 101.

\bibitem{trigener}
R. Bizzarri \etal , Nucl. Instr. Meth. {\bf A 283} (1989) 799.

\bibitem{l3tot}
L3 Collab., M. Acciarri \etal , \PL {\bf B 519} (2001) 33.

\end{thebibliography}

\newpage
\typeout{   }     
\typeout{Using author list for paper 281 -  }
\typeout{$Modified: Jul 15 2001 by smele $}
\typeout{!!!!  This should only be used with document option a4p!!!!}
\typeout{   }
%
%
%
%
%
%

\newcount\tutecount  \tutecount=0
\def\tutenum#1{\global\advance\tutecount by 1 \xdef#1{\the\tutecount}}
\def\tute#1{$^{#1}$}
\tutenum\aachen            
\tutenum\nikhef            
\tutenum\mich              
\tutenum\lapp              
\tutenum\basel             
\tutenum\lsu               
\tutenum\beijing           
\tutenum\bologna           
\tutenum\tata              
\tutenum\ne                
\tutenum\bucharest         
\tutenum\budapest          
\tutenum\mit               
\tutenum\panjab            
\tutenum\debrecen          
\tutenum\dublin            
\tutenum\florence          
\tutenum\cern              
\tutenum\wl                
\tutenum\geneva            
\tutenum\hefei             
\tutenum\lausanne          
\tutenum\lyon              
\tutenum\madrid            
\tutenum\florida           
\tutenum\milan             
\tutenum\moscow            
\tutenum\naples            
\tutenum\cyprus            
\tutenum\nymegen           
\tutenum\caltech           
\tutenum\perugia           
\tutenum\peters            
\tutenum\cmu               
\tutenum\potenza           
\tutenum\prince            
\tutenum\riverside         
\tutenum\rome              
\tutenum\salerno           
\tutenum\ucsd              
\tutenum\sofia             
\tutenum\korea             
\tutenum\purdue            
\tutenum\psinst            
\tutenum\zeuthen           
\tutenum\eth               
\tutenum\hamburg           
\tutenum\taiwan            
\tutenum\tsinghua          

{
\parskip=0pt
\noindent
{\bf The L3 Collaboration:}
\ifx\selectfont\undefined
 \baselineskip=10.8pt
 \baselineskip\baselinestretch\baselineskip
 \normalbaselineskip\baselineskip
 \ixpt
\else
 \fontsize{9}{10.8pt}\selectfont
\fi
\medskip
\tolerance=10000
\hbadness=5000
\raggedright
\hsize=162truemm\hoffset=0mm
\def\r{\rlap,}
\noindent

P.Achard\r\tute\geneva\ 
O.Adriani\r\tute{\florence}\ 
M.Aguilar-Benitez\r\tute\madrid\ 
J.Alcaraz\r\tute{\madrid}\ 
G.Alemanni\r\tute\lausanne\
J.Allaby\r\tute\cern\
A.Aloisio\r\tute\naples\ 
M.G.Alviggi\r\tute\naples\
H.Anderhub\r\tute\eth\ 
V.P.Andreev\r\tute{\lsu,\peters}\
% ...
\newpage
\begin{list}{A}{\itemsep=0pt plus 0pt minus 0pt\parsep=0pt plus 0pt minus 0pt
                \topsep=0pt plus 0pt minus 0pt}
\item[\aachen]
 III. Physikalisches Institut, RWTH, D-52056 Aachen, Germany$^{\S}$
\item[\nikhef] National Institute for High Energy Physics, NIKHEF, 
     and University of Amsterdam, NL-1009 DB Amsterdam, The Netherlands
\item[\mich] University of Michigan, Ann Arbor, MI 48109, USA
\item[\lapp] Laboratoire d'Annecy-le-Vieux de Physique des Particules, 
     LAPP,IN2P3-CNRS, BP 110, F-74941 Annecy-le-Vieux CEDEX, France
\item[\basel] Institute of Physics, University of Basel, CH-4056 Basel,
     Switzerland
\item[\lsu] Louisiana State University, Baton Rouge, LA 70803, USA
\item[\beijing] Institute of High Energy Physics, IHEP, 
  100039 Beijing, China$^{\triangle}$ 
\item[\bologna] University of Bologna and INFN-Sezione di Bologna, 
     I-40126 Bologna, Italy
\item[\tata] Tata Institute of Fundamental Research, Mumbai (Bombay) 400 005, India
\item[\ne] Northeastern University, Boston, MA 02115, USA
\item[\bucharest] Institute of Atomic Physics and University of Bucharest,
     R-76900 Bucharest, Romania
\item[\budapest] Central Research Institute for Physics of the 
     Hungarian Academy of Sciences, H-1525 Budapest 114, Hungary$^{\ddag}$
\item[\mit] Massachusetts Institute of Technology, Cambridge, MA 02139, USA
\item[\panjab] Panjab University, Chandigarh 160 014, India
\item[\debrecen] KLTE-ATOMKI, H-4010 Debrecen, Hungary$^\P$
\item[\dublin] Department of Experimental Physics,
  University College Dublin, Belfield, Dublin 4, Ireland
\item[\florence] INFN Sezione di Firenze and University of Florence, 
     I-50125 Florence, Italy
\item[\cern] European Laboratory for Particle Physics, CERN, 
     CH-1211 Geneva 23, Switzerland
\item[\wl] World Laboratory, FBLJA  Project, CH-1211 Geneva 23, Switzerland
\item[\geneva] University of Geneva, CH-1211 Geneva 4, Switzerland
\item[\hefei] Chinese University of Science and Technology, USTC,
      Hefei, Anhui 230 029, China$^{\triangle}$
\item[\lausanne] University of Lausanne, CH-1015 Lausanne, Switzerland
\item[\lyon] Institut de Physique Nucl\'eaire de Lyon, 
     IN2P3-CNRS,Universit\'e Claude Bernard, 
     F-69622 Villeurbanne, France
\item[\madrid] Centro de Investigaciones Energ{\'e}ticas, 
     Medioambientales y Tecnol\'ogicas, CIEMAT, E-28040 Madrid,
     Spain${\flat}$ 
\item[\florida] Florida Institute of Technology, Melbourne, FL 32901, USA
\item[\milan] INFN-Sezione di Milano, I-20133 Milan, Italy
\item[\moscow] Institute of Theoretical and Experimental Physics, ITEP, 
     Moscow, Russia
\item[\naples] INFN-Sezione di Napoli and University of Naples, 
     I-80125 Naples, Italy
\item[\cyprus] Department of Physics, University of Cyprus,
     Nicosia, Cyprus
\item[\nymegen] University of Nijmegen and NIKHEF, 
     NL-6525 ED Nijmegen, The Netherlands
\item[\caltech] California Institute of Technology, Pasadena, CA 91125, USA
\item[\perugia] INFN-Sezione di Perugia and Universit\`a Degli 
     Studi di Perugia, I-06100 Perugia, Italy   
\item[\peters] Nuclear Physics Institute, St. Petersburg, Russia
\item[\cmu] Carnegie Mellon University, Pittsburgh, PA 15213, USA
\item[\potenza] INFN-Sezione di Napoli and University of Potenza, 
     I-85100 Potenza, Italy
\item[\prince] Princeton University, Princeton, NJ 08544, USA
\item[\riverside] University of Californa, Riverside, CA 92521, USA
\item[\rome] INFN-Sezione di Roma and University of Rome, ``La Sapienza",
     I-00185 Rome, Italy
\item[\salerno] University and INFN, Salerno, I-84100 Salerno, Italy
\item[\ucsd] University of California, San Diego, CA 92093, USA
\item[\sofia] Bulgarian Academy of Sciences, Central Lab.~of 
     Mechatronics and Instrumentation, BU-1113 Sofia, Bulgaria
\item[\korea]  The Center for High Energy Physics, 
     Kyungpook National University, 702-701 Taegu, Republic of Korea
\item[\purdue] Purdue University, West Lafayette, IN 47907, USA
\item[\psinst] Paul Scherrer Institut, PSI, CH-5232 Villigen, Switzerland
\item[\zeuthen] DESY, D-15738 Zeuthen, Germany
\item[\eth] Eidgen\"ossische Technische Hochschule, ETH Z\"urich,
     CH-8093 Z\"urich, Switzerland
\item[\hamburg] University of Hamburg, D-22761 Hamburg, Germany
\item[\taiwan] National Central University, Chung-Li, Taiwan, China
\item[\tsinghua] Department of Physics, National Tsing Hua University,
      Taiwan, China
\item[\S]  Supported by the German Bundesministerium 
        f\"ur Bildung, Wissenschaft, Forschung und Technologie.
\item[\ddag] Supported by the Hungarian OTKA fund under contract
numbers T019181, F023259 and T037350.
\item[\P] Also supported by the Hungarian OTKA fund under contract
  number T026178.
\item[$\flat$] Supported also by the Comisi\'on Interministerial de Ciencia y 
        Tecnolog{\'\i}a.
\item[$\sharp$] Also supported by CONICET and Universidad Nacional de La Plata,
        CC 67, 1900 La Plata, Argentina.
\item[$\triangle$] Supported by the National Natural Science
  Foundation of China.
\end{list}
}
\vfill


\newpage



\begin{table}
  \begin{center}

\begin{tabular}{|c|c|| c| c |c| c|}    
        \hline
$p_t$& $\langle p_t\rangle$&  Number of $\Lambda$ & Mass of $\Lambda$
& Efficiency& $d\sigma / d p_t$ \\ 
(\GeV) & (\GeV) & &(\MeV)  & (\%) & (pb/\GeV) \\\hline\hline
   0.4$-$0.6  
 & 0.50 
 & $3412 \pm 71$ 
 & $1116.0 \pm 0.1$		
 & $10.2 \pm 0.1$ 
 & 273  $\pm$    6  $\pm$   36\\\hline
   0.6$-$0.8  
 & 0.69 
 & $4408 \pm 85$ 
 & $1115.9 \pm 0.1$		
 & $13.7 \pm 0.1$ 
 & 264  $\pm$    5  $\pm$   29\\\hline
   0.8$-$1.0  
 & 0.89 
 & $3420 \pm 81$ 
 & $1116.1 \pm 0.1$		
 & $15.3 \pm 0.2$
 & 183  $\pm$    4  $\pm$   15\\\hline
   1.0$-$1.3  
 & 1.12 
 & $3201 \pm 87$ 
 & $1115.7 \pm 0.1$
 & $16.8 \pm 0.2$ 
 & 104  $\pm$    3  $\pm$   \phantom{0}7\\\hline
   1.3$-$1.6  
 & 1.43 
 & $1222 \pm 55$ 
 & $1115.5 \pm 0.1$		
 & $17.7 \pm 0.4$ 
 & \phantom{0}38  $\pm$    2  $\pm$    \phantom{0}3\\\hline
   1.6$-$2.0  
 & 1.77 
 & $\phantom{0}578 \pm 41$	 
 & $1115.0 \pm 0.2$
 & $15.9 \pm 0.6$ 
 & \phantom{0}15  $\pm$    1  $\pm$    \phantom{0}1\\\hline
 2.0$-$2.5  
 & 2.21 
 & $\phantom{0} 292 \pm 25 $	 
 & $1115.9 \pm 0.3$
 & $17.2\pm 1.3$ 
 & \phantom{00}6  $\pm$    1  $\pm$    \phantom{0}1  \\
\hline
\end{tabular}

\caption{The number of $\Lambda$ baryons estimated by the fit, together with the $\Lambda$ mass, 
the overall efficiency and the corresponding differential cross section as a function of 
$p_t$ for $|\eta| < 1.2$. The first uncertainty on the cross section is statistical and 
the second systematic.}
\label{tab1}
\end{center}
\end{table}


\begin{table}
  \begin{center}

\begin{tabular}{|c|| c| c|c|}    
        \hline
$|\eta|$& Number of $\Lambda$  & 
Efficiency& $d\sigma / d |\eta|$ \\ 
  &  & (\%) & (pb) \\\hline \hline
 0.0$-$0.3        &  
 2953   $\pm$ 72  & 
 13.8   $\pm$ 0.2 & 
 59     $\pm$  1 $\pm$ 3  \\ \hline

 0.3$-$0.6        &
 2742   $\pm$ 63  &
 13.4   $\pm$ 0.2 &
 56     $\pm$  1 $\pm$ 3  \\ \hline
 0.6$-$0.9        &  
 2904   $\pm$ 70  &
 13.0   $\pm$ 0.2 &
 61     $\pm$  1 $\pm$ 4  \\ \hline
 0.9$-$1.2        &
 2774   $\pm$ 89  &
 11.1   $\pm$ 0.1  &
 68     $\pm$  2 $\pm$ 8  \\
    \hline
\end{tabular}
\caption{The number of $\Lambda$ baryons estimated by the fit, together with 
the overall efficiency and the corresponding differential cross section as a function of
pseudorapidity for $0.4 \GeV < p_t \leq  1 \GeV$. The first uncertainty on the cross section is 
statistical and the second systematic.}
\label{tab2}
\end{center}
\end{table}


\begin{table}
  \begin{center}

\begin{tabular}{|c|| c| c|c|}    
        \hline
$|\eta|$& Number of $\Lambda$  & 
Efficiency& $d\sigma / d |\eta|$ \\ 
  & & (\%) & (pb) \\\hline \hline

 0.0$-$0.3   & 
 1458   $\pm$ 60   & 
 18.8   $\pm$ 0.5  &  
 21     $\pm$ 1  $\pm$ 1 \\\hline
 0.3$-$0.6         & 
 1411   $\pm$ 62   & 
 18.2   $\pm$ 0.4  &  
 21     $\pm$ 1  $\pm$ 1 \\\hline
 0.6$-$0.9         & 
 1480   $\pm$ 62   &       
 19.3   $\pm$ 0.5  &  
 21     $\pm$ 1  $\pm$ 1 \\\hline
 0.9$-$1.2   & 
 1007   $\pm$ 63 & 
 14.3   $\pm$ 0.4  &  
 19     $\pm$ 1  $\pm$ 3  \\\hline
\end{tabular}

\caption{The number of $\Lambda$ baryons estimated by the fit, together with 
the overall efficiency and the corresponding differential cross section as a function of
pseudorapidity for $1 \GeV < p_t < 2.5 \GeV$. The first uncertainty on the cross section 
is statistical and the second systematic.}
\label{tab3}
\end{center}
\end{table}


\begin{table}
  \begin{center}

\begin{tabular}{|c|c| c| c |c | c| }    
        \hline
$p_t$ & Background & Scales and & Monte Carlo & Trigger & Total\\
(\GeV) & subtraction (\%) & resolutions (\%) & modelling (\%) & efficiency (\%) &(\%) \\ \hline
0.4$-$0.6 &           12.1  & 4.2 & 1.4  & 2.0 &            13.0 \\ 
0.6$-$0.8 & \phantom{0}9.8  & 4.2 & 1.6  & 2.0 &            11.0 \\ 
0.8$-$1.0 & \phantom{0}6.6  & 4.2 & 1.9  & 2.0 &  \phantom{0}8.3 \\ 
1.0$-$1.3 & \phantom{0}5.3  & 4.2 & 2.1  & 2.0 &  \phantom{0}7.3  \\ 
1.3$-$1.6 & \phantom{0}5.8  & 4.2 & 2.4  & 2.0 &  \phantom{0}7.8  \\ 
1.6$-$2.0 & \phantom{0}6.4  & 4.2 & 2.8  & 2.0 &  \phantom{0}8.4 \\ 
2.0$-$2.5 &           19.4  & 4.2 & 4.8  & 2.0 &            20.5 \\ 
\hline
\end{tabular}

\caption{Systematic uncertainty on the cross section of the $\epem \ra \epem \Lambda  \rm X $ and 
$\epem \ra \epem \overline{\Lambda} \rm X $ processes due to background subtraction, 
scales and resolution uncertainties, Monte Carlo modelling and trigger efficiency. The total systematic uncertainty 
is the quadratic sum of the different contributions.}
\label{tabsyst}
\end{center}
\end{table}



\newpage


\newpage


\newpage


\newpage


\newpage


\newpage

\end{document}